\documentclass{ws-procs9x6-cpt16}
\begin{document}

\newcommand{\refeq}[1]{(\ref{#1})}
\mathchardef\mhyphen="2D
\def\etal {{\it et al.}}

\title{ Constraining Anomalous Forces with Pseudoscalar and Axial Couplings Employing a Spin-Independent Analysis}

\author{S.\ Aldaihan,$^1$ W.M.\ Snow,$^{1}$ D.E.\ Krause,$^{2,3}$  J.C.\ Long$^1$, and E.\ Fischbach$^3$}

\address{$^1$Physics Department, Indiana University, Bloomington, Indiana 47408, U.S.A}

\address{$^2$Physics Department, Wabash College, Crawfordsville, Indiana 47933, U.S.A}

\address{$^3$Department of Physics and Astronomy, Purdue University, West Lafayette, Indiana 47907, U.S.A}
\begin{abstract}
Present laboratory limits on the coupling strength of anomalous pseudoscalar and axial interactions are many orders of magnitude weaker than their scalar and vector analogs. Here we investigate two mechanisms which can circumvent this suppression and thereby lead to improved limits. 
\end{abstract}

\bodymatter

\vspace{12pt}


Exotic long-range interactions can be generated by many possible sources beyond the Standard Model. Very stringent constraints exist on spin-independent Yukawa interactions arising from ultralight scalar or vector bosons. For such interactions, the present constraint on the respective dimensionless coupling constant is  $\sim 10^{-40}$ if the range of the force is $\sim 1$~mm.\cite{heckel2007} In contrast, the experimental limits on interactions with pseudoscalar and axial couplings are several orders of magnitude weaker than limits on spin-independent Yukawa interactions. For example, within the same range, the constraint on the pseudoscalar coupling to electrons is $\sim10^{-16}$.\cite{heckel2015} The laboratory limits on couplings to nucleons are weaker still. Similar argument follow for the axial coupling to nucleons, which is $\sim10^{-13}$.\cite{PSI}  Stringent limits on  axial couplings to electrons exist, but these apply to very long length scales.\cite{heckel2013} Further, limits on pseudoscalar and axial couplings are  much weaker at the atomic scale where, to date, no direct experiments exist, and most limits are inferred from other techniques as can be deduced from Ref.~\refcite{ledbetter}. Among the reasons for differing sensitivities is the fact that pseudoscalar and axial bosons necessarily couple at lowest order to matter through spin-dependent interactions, so polarization of the test samples is needed in order to obtain a coherent effect from such interactions.   This introduces practical difficulties such as only a small fraction of the sample is polarized and comparatively large competing magnetic forces need to be eliminated. Further, most spin-dependent interactions have parity-odd vertices.  Therefore, in order for parity to be conserved, the boson must be emitted with nonzero angular momentum relative to the initial and final fermions. This angular momentum has a ``memory" of order $(1/m r)$, where $m$ is the mass of the fermion and $r$ is the distance between the interacting fermions. Due to the disparity between the limits from spin-dependent and spin-independent experiments, we raise the question whether new limits on spin-dependent couplings can be inferred from the analysis of existing spin-independent data. Here we exploit two mechanisms by which this can be achieved. 


In the first approach, we note that from the helicity of spin-1/2 particles, a spin-independent force is generated by the exchange of 2-pseudoscalars\cite{2-pseudoscalar} and 2-axial vector bosons between two spin-1/2 particles.  By calculating the very long range contribution of such interactions, we found the leading potentials to be
\begin{equation}
V_\mathrm{P\mhyphen P}(r)=-\frac{g^4_\mathrm{P}}{32 \pi^3  m^2}\frac{\mu K_\mathrm{1}(2\mu r)}{r^2},
\label{aba:eq1}
\end{equation}
\begin{equation}
V_\mathrm{S\mhyphen P}(r)=  \frac{3 \: g^2_\mathrm{S}g^2_\mathrm{P}}{64 \pi^2 m }   \frac{e^{-2 \mu r}}{ r^2},
\label{aba:eq2}
\end{equation}
where $V_\mathrm{P\mhyphen P}$ is the potential due to 2-pseudoscalar exchange and $V_\mathrm{S\mhyphen P}$ is due to the exchange of 2-bosons between scalar and pseudoscalar vertices. Here $\mu$ is the mass of the exchanged boson, $g_\mathrm{S}$ and $g_\mathrm{P}$ are the scalar and pseudoscalar coupling constants, respectively, and $K_1(x)$ is the modified Bessel function.  Eq.~\refeq{aba:eq1} agrees with previous work.\cite{Drell,Ferrer} Evaluation of the functional forms for 2-axial exchange and 2-boson exchange between vector and axial vertices is in progress.
A 2012 data set from the short range gravity experiment at Indiana University \cite{long} has been analyzed for an interaction of the form in Eq.~\refeq{aba:eq1}.  Preliminary results suggest an improved constraint on $(g^N_\mathrm{P})^2$, the pseudoscalar coupling to nucleons, in the range between $40$ and $200~\mu$m of about a factor of 5 compared to previous limits.\cite{Klimchitskaya}  We are currently extending the same analysis to $(g^e_\mathrm{P})^2$ and $g_\mathrm{S} g_\mathrm{P}$. 


The other process by which the spin dependence can be eliminated is by exchanging the pseudoscalar or the axial vector bosons between separate bound state systems, say two nuclei. Whereas averaging a spin-dependent interaction over the spin states of unpolarized fermions yields no coherent contribution, in the case of nuclei, it may be possible to find a combination of quantum numbers for orbital angular momentum $L$ and spin $S$  that give a nonzero contribution. For two nuclei, say $A$ and $B$, interacting via a spin-dependent interaction has an effective potential
\begin{equation}
V_ {\mathrm{A,B}}(r) \sim  g^2 \langle Q_\mathrm{A}\rangle  \langle Q_\mathrm{B} \rangle  \frac{e^{- \mu r}}{4\pi r},
\label{aba:eq3}
\end{equation}
where $\langle Q_\mathrm{A/B}\rangle$ is the matrix element of the nucleons inside each nucleus and $g$ is the anomalous spin-dependent coupling we wish to constrain. Another advantage of such interaction, first identified  in Ref.~\refcite{krause}, is that all the suppression factors associated with parity-violating vertices are absorbed by the mean field of the nuclei since, within the nucleus, $(1/m r) \sim 1$. The theoretical evaluation of such interaction requires knowledge of matrix elements in nuclei exploiting all possible interactions, both parity-even and odd, that eliminate the spin-dependence. Such analysis will significantly improve existing limits on spin dependent couplings, including P- and T-odd and axion motivated, at both long as well as short distance scales provided that there exist experiments searching for long range forces between bound state systems (i.e, nuclei, atoms, molecules).


\begin{thebibliography}{xx}
\bibitem{heckel2007}
D.J.\ Kapner, \etal, Phys.\ Rev.\ Lett.\ {\bf 98}, 021101 (2007).
\bibitem{heckel2015} 
W.A.\ Terrano \etal, Phys.\ Rev.\ Lett.\ {\bf 115},  201801 (2015).
\bibitem{PSI} 
F.M.\ Piegsa and G.\ Pignol, Phys.\ Rev.\ Lett.\ {\bf 108}, 181801 (2012).
\bibitem{heckel2013} 
B.R.\ Heckel \etal, Phys.\ Rev.\ Lett.\ {\bf111}, 15802 (2013).
\bibitem{ledbetter} 
M.P.\ Ledbetter \etal, Phys.\ Rev.\ Lett.\ {\bf 110}, 040402 (2013); N.\ F.\ Ramsey, Physica (Amsterdam) {\bf 96A}, 285 (1979).
\bibitem{2-pseudoscalar}
E.\ Fischbach and D.E.\ Krause, Phys.\ Rev.\ Lett.\ {\bf 82}, 4753 (1999); E.\ Fischbach and D.E.\ Krause, Phys.\ Rev.\ Lett.\ {\bf 83}, 3593 (1999); E.G.\ Adelberger \etal, Phys. Rev. D {\bf 68}, 062002 (2003).\bibitem{Drell} 
S.D.\ Drell and K.\ Huang.\ Phys.\ Rev.\ {\bf 91}, 1527 (1953). 
\bibitem{Ferrer} 
F.\ Ferrer and M.\ Nowakowski, Phys.\ Rev.\ D {\bf 59}, 075009 (1999).
\bibitem{long} 
J.C.\ Long and V.\ Alan Kostelecky, Phys.\ Rev.\ D {\bf 91}, 092003 (2015).
\bibitem{Klimchitskaya} 
G.\ Klimchitskaya and V.M.\ Mostepanenko, Eur.\ Phys.\ J.\ C, {\bf 75},  164  (2015). 
\bibitem{krause} 
D.E.\ Krause \etal, in {\em Perspectives in Neutrinos, Atomic Physics, and Gravitation}, Editions Fronti\'{e}rs, Gif-sur-Yvette Cedex, France, pp.~455--463 (1993).
\end{thebibliography}
\end{document}